\documentclass[12pt]{article}
\usepackage{float}
\usepackage{amsfonts}
\usepackage{amsthm}
\usepackage{eurosym}
\usepackage{graphicx}
\usepackage[utf8]{inputenc}
\usepackage[T1]{fontenc}
\usepackage{indentfirst}
\usepackage[margin=0.6in,nomarginpar]{geometry}
\usepackage[final]{hyperref}
\usepackage{amsmath}
\usepackage{hyperref}
\usepackage{cite}
\usepackage{subcaption}
\usepackage{caption}
\usepackage{amssymb}
\usepackage{multirow}
\usepackage[table]{xcolor}
\usepackage{orcidlink}
\usepackage{authblk}
\usepackage{float}
\hypersetup{
colorlinks=true,
linkcolor=blue,
citecolor=blue,
filecolor=magenta,
urlcolor=blue
}

\begin{document}

\title{Dunkl-Klein-Gordon Equation in Higher Dimensions}
\author{B. Hamil \orcidlink{0000-0002-7043-6104} \thanks{%
hamilbilel@gmail.com/bilel.hamil@umc.edu.dz (Corresponding author)} \\
Laboratoire de Physique Mathématique et Subatomique, \\
Faculté des Sciences Exactes, Université Constantine 1 Frères Mentouri, Constantine, Algeria.\\
 \and B. C. L\"{u}tf\"{u}o\u{g}lu %
\orcidlink{0000-0001-6467-5005} \thanks{%
bekir.lutfuoglu@uhk.cz } \\
Department of Physics, Faculty of Science, University of Hradec Kralove, \\
Rokitanskeho 62/26, Hradec Kralove, 500 03, Czech Republic. \and M. Merad 
\orcidlink{0000-0001-7547-6933} \thanks{
meradm@gmail.com} \\
Laboratoire de systèmes dynamiques et contr\^{o}le (L.S.D.C), \\
Département des sciences de la matière, Faculté des Sciences Exactes\\et SNV, Université de Oum-El-Bouaghi, 04000, Oum El Bouaghi, Algeria.}
\date{}
\maketitle

\begin{abstract}
In this study, we replace the standard partial derivatives in the Klein-Gordon equation with Dunkl derivatives and obtain exact analytical solutions for the eigenvalues and eigenfunctions of the Dunkl-Klein-Gordon equation in higher dimensions. We apply this formalism to two key quantum mechanical systems: the d-dimensional harmonic oscillator and the Coulomb potential. First, we introduce Dunkl quantum mechanics in d-dimensional polar coordinates, followed by an analysis of the d-dimensional Dunkl-Klein-Gordon oscillator. Subsequently, we derive the energy spectrum and eigenfunctions, which are expressed using confluent hypergeometric functions. Furthermore, we examine the impact of the Dunkl formalism on both the eigenvalues and eigenfunctions. In the second case, we explore both the bound-state solutions and scattering scenarios of the Dunkl-Klein-Gordon equation with the Coulomb potential. The bound-state solutions are represented in terms of confluent hypergeometric functions, while the scattering states enable us to compute the particle creation density and probability using the Bogoliubov transformation method.

\end{abstract}

\section{Introduction}
For over a century, physicists have pursued novel methodologies and developed new formalisms to elucidate the behavior of quantum mechanical systems. One such formalism is the Dunkl formalism, which proposes the use of the Dunkl derivative in place of the traditional derivative. This formalism demonstrates that the solutions of the systems under consideration can be classified according to the reflection symmetries of the particles. The Dunkl derivatives were initially introduced by the mathematician Charles Dunkl in 1989 as a combination of differential and difference operators associated with a finite reflection group in the context of mathematics, as follows \cite{Dunkl1989}: 
\begin{eqnarray} 
D_x&=&\frac{\partial }{\partial x}+\frac{\mu} {x}\left( \mathbf{1}-R\right). 
\end{eqnarray} 
Here, $\mu$ is a constant and $R$ is the reflection operator: \begin{eqnarray} 
Rf(x)=f(-x). 
\end{eqnarray} 
Interestingly, a similar derivative known as the Yang derivative, \begin{eqnarray} 
D_Y&=&\frac{\partial }{\partial x}-\frac{\mu } {x}R, 
\end{eqnarray} 
was proposed in a series of papers by Wigner and Yang in the mid-20th century, following a highly original conceptual discussion of quantum mechanics \cite{Wigner1950, Yang1951}. The resemblance between the two operators prompted a group of physicists to utilize the Dunkl operator in deformed algebras \cite{Plyushchay1994, Plyushchay1996, Plyushchay1997}, and in Calogero \cite{Chakra1994}, Calogero-Sutherland \cite{Lapointe1996, Kakei1996}, and Calogero-Sutherland-Moser models \cite{Hikami1996}, as the Yang derivative had previously been a valuable tool in introducing color degrees of freedom and quantum chromodynamics \cite{Green1953, Greenberg1964}. In the following years, interest in the use of the Dunkl operator in physics remained limited, despite the significant contributions of Plyushchay and his colleagues \cite{Gamboa1999, Plyushchay2000, Klishevich2001, Hovarthy2004, Rodrigues2009, Horvathy2010, Bie2012, Correa2014}. This perspective was reoriented in 2013, following the publication of a series of papers by Genest et al. \cite{Genest20131, Genest20132, Genest20133, Genest20141, Genest20142}, who applied the Dunkl operator to the Schrödinger equation to investigate two- and three-dimensional isotropic and anisotropic oscillators. Subsequently, Dunkl-Schrödinger equation solutions with the Coulomb interaction were analyzed in two, three, and arbitrary dimensions, respectively \cite{Vincent3, Ghazouani2020, Ghazouani2021}. Meanwhile, Salazar-Ramírez et al. obtained coherent state solutions of the Dunkl oscillator and Dunkl-Coulomb problems using the su(1,1) Lie algebra in two dimensions \cite{Salazar2017, Salazar2018}. It is also noteworthy that mathematical generalizations and algebraic techniques have been employed to explore the properties of various Dunkl oscillator models in two- and three-dimensional curved \cite{Najafizade20221, Najafizade20222, Najafizade20223, Ballesteros2024} and noncommutative phase spaces \cite{Samira2023}.

Recent advances in relativistic quantum mechanics have led to the integration of the Dunkl operator into various fundamental equations, resulting in significant developments such as the Dunkl-Dirac (DD) oscillator \cite{Sargol2018, Mota20181}, the Dunkl-Klein-Gordon (DKG) oscillator \cite{Mota20212, Bilel20221}, and the Dunkl-Duffin-Kemmer-Petiau (DDKP) oscillator \cite{Merad2021, Merad2022} in both one and two dimensions. Further research demonstrated that algebraic methods could be used to determine the energy spectrum of the one-dimensional DD oscillator \cite{Ojeda2020} and highlighted a connection to the Anti-Jaynes-Cummings Model \cite{Bilel20222}. Solutions for Landau levels in the DKG oscillator, coupled with an external magnetic field in two dimensions, were obtained using $su(1,1)$ Lie algebra techniques and analytical methods \cite{Mota20211}. The DKG oscillator in various dimensions was also analyzed using the path integral formalism, leading to the derivation of exact propagators, energy eigenvalues, and wave functions \cite{Benzair2024}. Additionally, closed-form solutions for the DKG equation involving inverse power-law interactions were provided \cite{Schulze20243}. A subsequent study on the DDKP equation, addressing step potential scattering and the Ramsauer–Townsend effect, was presented in \cite{Askari2023}.

In recent years, particularly over the last three, the Dunkl formalism has found widespread application across various branches of physics, including quantum statistics \cite{Hassan2021, Fateh20231, Fateh20232}, condensed matter \cite{Hocine20241, Hocine20242}, solid-state physics \cite{Bilel20223}, quantum optics \cite{Chung2024}, and other areas of theoretical physics \cite{Junker2023, Quesne2023, Quesne2024, Schulze20241, Schulze20242, Mota20241, Mota20242, Junker, Bouguerne2024, Khantoul20241, Benchikha20241, Khantoul20242, Bouguerne20241}. To explore the most generalized solutions, researchers have proposed extensions of the Dunkl formalism, including various forms of Dunkl derivatives \cite{Dong2021, Halberg2022, Samira2022, Mota20221, Dong2023, Rouabhia2023} and solutions in arbitrary dimensions \cite{Ghazouani2021, Hamil2024}. In this contribution, we aim to address an important gap in the literature by developing the DKG equation in d-dimensions and investigating its oscillator solution. Additionally, we will explore the DKG equation under the Coulomb potential to provide a comprehensive understanding of both bound-state solutions and scattering cases. 

We form the manuscript as follows: In section \ref{sec2}, we introduce d-dimensional polar coordinates and adapt the Dunkl formalism along with the Klein-Gordon equation accordingly. We also discuss the angular part solution. Section \ref{sec3} delves into the solution of the radial equation for harmonic oscillator potentials. In section \ref{sec4}, solutions for the Coulomb potential are derived for both bound-state and scattering cases. Finally, the manuscript concludes with a brief summary.

\section{Dunkl-Klein-Gordon in d-dimensions: Polar coordinates}

\label{sec2}

In this section, we analyze the DKG equation in d-dimensions using hyperspherical coordinates. We begin by considering the DKG equation in arbitrary d-dimensions,
\begin{equation}
\left[ E^{2}+D_{j}^{2}-m^{2}\right] \Psi \left( \mathbf{x}\right) =0, \label{kg01}
\end{equation}%
where $\mathbf{x}$ is a d-dimensional position vector with the hyperspherical Cartesian components $x_{1}$; $x_{2}$; $\cdots $; $x_{d}$ given as:
\begin{equation} 
\left. 
\begin{array}{c}
x_{1}=r\cos \theta _{1}\sin \theta _{2}\sin \theta _{3}\cdots \sin \theta
_{d-1} \\ 
x_{2}=r\sin \theta _{1}\sin \theta _{2}\sin \theta _{3}\cdots \sin \theta
_{d-1} \\ 
x_{3}=r\cos \theta _{2}\sin \theta _{3}\sin \theta _{4}\cdots \sin \theta
_{d-1} \\ 
\vdots \\ 
x_{j}=r\cos \theta _{j-1}\sin \theta _{j}\sin \theta _{j+1}\cdots \sin
\theta _{d-1} \\ 
\vdots \text{\ } \\ 
x_{d}=r\cos \theta _{d-1}\text{ \ \ \ \ \ \ \ \ \ \ \ \ \ \ \ \ \ \ \ \ \ \
\ \ \ \ \ \ \ \ }
\end{array}%
\right.  \label{e5},  
\end{equation}%
where $j\in \left[ 3,d-1\right] $ and the range of the variables is $r\in \left] 0,+\infty \right[ ,$ $0\leq \theta _{1}\leq 2\pi $ and $0\leq \theta_{k}\leq \pi ,$ with $k\in \left[ 2,d-1\right] $. Summing the squares of Eqs. (\ref{e5}) yields%
\begin{equation}
\sum_{j=1}^{d}x_{j}^{2}=r^{2}.
\end{equation}%
In two dimensions ($d=2$), this describes the transformation from  polar coordinates ($r$, $\theta$) to Cartesian coordinates, where $x_{1}\equiv x$ and $x_{2} \equiv y$. In three dimensions ($d = 3$), it represents the transformation from spherical coordinates ($r$, $\theta_{1}$, $\theta_{2}$) to Cartesian coordinates.

In hyperspherical coordinates, the Laplacian operator is written as follows:
\begin{eqnarray}
\Delta &=&\frac{\partial ^{2}}{\partial r^{2}}+\frac{d-1}{r}\frac{\partial }{\partial r}+\frac{1}{r^{2}}\sum_{j=1}^{d-2}\frac{1}{\sin ^{2}\theta _{j+1}\sin ^{2}\theta _{j+2}\cdots \sin ^{2}\theta _{d-1}}\left\{ \frac{ \partial ^{2}}{\partial \theta _{j}^{2}}+\left( j-1\right) \tan \theta _{j} \frac{\partial }{\partial \theta _{j}}\right\}  \notag \\
&&+\frac{1}{r^{2}}\left\{ \frac{1}{\sin ^{d-2}\theta _{d-1}}\frac{\partial }{\partial \theta _{d-1}}\sin ^{d-2}\theta _{d-1}\frac{\partial }{\partial \theta _{d-1}}\right\} ,\label{107}
\end{eqnarray}
while
\begin{equation}
\prod\limits_{j=1}^{d}dx_{j}=r^{d-1}dr\prod\limits_{j=1}^{d-1}\left( \sin \theta _{j}\right) ^{j-1}d\theta _{j}.
\end{equation}%
The substitution of Eqs. (\ref{e5}) and (\ref{107}) into Eq.~ (\ref{kg01}) yields the d-dimensional DKG equation in hyperspherical  coordinates:
\begin{equation}
\left[ \mathcal{A}_{r}+\frac{\mathcal{J}_{\theta _{1}}}{r^{2}\sin ^{2}\theta
_{2}\sin ^{2}\theta _{3}\cdots\sin ^{2}\theta _{d-1}}+\frac{\mathcal{J}_{\theta
_{2}}}{\sin ^{2}\theta _{3}\cdots\sin ^{2}\theta _{d-1}}+\cdots +\frac{1}{r^{2}}%
\mathcal{J}_{\theta _{d-1}}\right] \Psi \left( \mathbf{x}\right) =0,\label{810}
\end{equation}%
where 
\begin{equation}
\mathcal{A}_{r}=\frac{\partial ^{2}}{\partial r^{2}}+\frac{d-1+2\left( \mu
_{1}+\mu _{2}+\mu _{3}+\cdots+\mu _{d}\right) }{r}\frac{\partial }{\partial r}%
+E ^{2}-m^{2}.
\end{equation}
Here, the angular operators $\mathcal{J}_{\theta_{1}},\mathcal{J}_{\theta_{2}},\cdots ,\mathcal{J}_{\theta _{d-1}}$, are expressed in  hyperspherical coordinates as:
\begin{equation}
\left. 
\begin{array}{c}
\mathcal{J}_{\theta _{1}}=-\frac{\partial ^{2}}{\partial \theta _{1}^{2}}+%
\frac{2\left( \mu _{1}\tan \theta _{1}-\mu _{2}\cot \theta _{1}\right) }{%
r^{2}}\frac{\partial }{\partial \theta _{1}}+\frac{\mu _{1}}{\cos ^{2}\theta
_{1}}\left( 1-R_{1}\right) +\frac{\mu _{2}\left( 1-R_{2}\right) }{\sin
^{2}\theta _{1}} \\ 
\mathcal{J}_{\theta _{2}}=-\frac{\partial ^{2}}{\partial \theta _{2}^{2}}-%
\left[ \left( 1+2\left( \mu _{1}+\mu _{2}\right) \right) \cot \theta
_{2}-2\mu _{3}\tan \theta _{2}\right] \frac{\partial }{\partial \theta _{2}}+%
\frac{\mu _{3}\left( 1-R_{3}\right) }{\cos ^{2}\theta _{2}} \\ 
\mathcal{J}_{\theta _{3}}=-\frac{\partial ^{2}}{\partial \theta _{3}^{2}}-%
\left[ \left( 2+2\left( \mu _{1}+\mu _{2}+\mu _{3}\right) \right) \cot
\theta _{3}-2\mu _{4}\tan \theta _{3}\right] \frac{\partial }{\partial
\theta _{3}}+\frac{\mu _{4}\left( 1-R_{4}\right) }{\cos ^{2}\theta _{3}} \\ 
\vdots \\ 
\mathcal{J}_{\theta _{d-2}}=-\frac{\partial ^{2}}{\partial \theta _{d-2}^{2}}%
-\left[ \left( \left( d-3\right) +2\left( \mu _{1}+\mu _{2}+\cdots+\mu
_{d-2}\right) \right) \cot \theta _{d-2}-2\mu _{d-1}\tan \theta _{d-2}\right]
\frac{\partial }{\partial \theta _{d-2}}+\frac{\mu _{d-1}\left(
1-R_{d-1}\right) }{\cos ^{2}\theta _{d-2}}\text{\ } \\ 
\mathcal{J}_{\theta _{d-1}}=\frac{\partial ^{2}}{\partial \theta _{d-1}^{2}}+%
\left[ \left( \left( d-2\right) +2\left( \mu _{1}+\mu _{2}+\cdots+\mu
_{d-1}\right) \right) \cot \theta _{d-1}-2\mu _{d}\tan \theta _{d-1}\right] 
\frac{\partial }{\partial \theta _{d-1}}+\frac{\mu _{d}\left( 1-R_{d}\right) 
}{\cos ^{2}\theta _{d-1}}%
\end{array}%
\right. .
\end{equation}
These angular operators depend only on the angles $\theta _{1},\cdots,\theta _{d-1}$, and their eigenstates are functions of $\theta _{1},\cdots ,\theta_{d-1}$, denoted as $\Theta _{1}\left( \theta_{1}\right) ,\Theta _{2}\left( \theta _{2}\right) ,\cdots ,\Theta
_{d-1}\left( \theta _{d-1}\right) $. The action of the reflection operators is readily observed to be
\begin{equation}
\left. 
\begin{array}{c}
R_{1}f\left( r,\theta _{1},\theta _{2},\cdots ,\theta _{j},\cdots ,\theta
_{d-1}\right) =f\left( r,\pi -\theta _{1},\theta _{2},\cdots ,\theta
_{j},\cdots ,\theta _{d-1}\right) \\ 
R_{2}f\left( r,\theta _{1},\theta _{2},\cdots ,\theta _{j},\cdots ,\theta
_{d-1}\right) =f\left( r,-\theta _{1},\theta _{2},\cdots ,\theta _{j},\cdots
,\theta _{d-1}\right) \\ 
\vdots \\ 
R_{j}f\left( r,\theta _{1},\theta _{2},\cdots ,\theta _{j},\cdots ,\theta
_{d-1}\right) =f\left( r,\theta _{1},\theta _{2},\cdots ,\pi -\theta
_{j},\cdots ,\theta _{d-1}\right) \\ 
\vdots \\ 
R_{d}f\left( r,\theta _{1},\theta _{2},\cdots ,\theta _{j},\cdots ,\theta
_{d-1}\right) =f\left( r,\theta _{1},\theta _{2},\cdots ,\theta _{j},\cdots
,\pi -\theta _{d-1}\right)%
\end{array}%
\right. .
\end{equation}%
Let us now use a separable solution of the form
\begin{equation}
\psi \left( r,\theta _{1},\cdots ,\theta _{d-1}\right) =\mathcal{R}\left(
r\right) \Theta _{1}\left( \theta _{1}\right) \Theta _{2}\left( \theta
_{2}\right) \cdots \Theta _{d-1}\left( \theta _{d-1}\right) .
\end{equation}%
Substituting this into Eq.~ (\ref{810}) gives us one radial 
\begin{equation}
\left[ \frac{\partial ^{2}}{\partial r^{2}}+\frac{d-1+2\left( \mu _{1}+\mu_{2}+\mu _{3}+\cdots+\mu _{d}\right) }{r}\frac{\partial }{\partial r}+E ^{2}-m^{2}-\frac{\varpi ^{2}}{r^{2}}\right] \mathcal{R}\left( r\right) =E\mathcal{R}\left( r\right) ,
\end{equation}%
and d-1 angular equations
\begin{equation}
\left( \mathcal{J}_{\theta _{1}}+\lambda _{1}^{2}\right) \Theta _{1}\left(
\theta _{1}\right) =0,  \label{1a}
\end{equation}%
\begin{equation}
\left( \mathcal{J}_{\theta _{2}}+\frac{\lambda _{1}^{2}}{\sin ^{2}\theta _{2}%
}+\lambda _{2}^{2}\right) \Theta _{2}\left( \theta _{2}\right) =0,
\end{equation}%
\begin{equation*}
\vdots
\end{equation*}%
\begin{equation}
\left( \mathcal{J}_{\theta _{d-1}}+\frac{\lambda _{d-2}^{2}}{\sin ^{2}\theta
_{d-1}}+\varpi ^{2}\right) \Theta _{d-1}\left( \theta _{d-1}\right) =0.
\end{equation}%
where $\lambda _{1},\lambda _{2},\cdots,\lambda _{d-2}$ and $\varpi $ are the separation constants, which play a crucial role in connecting the various differential equations. The solutions for the angular component $\Theta_{1}\left( \theta _{1}\right) $ is characterized by the eigenvalues of two reflection operator $R_{1}$ and $R_{2}$, denoted by $s_{1}=\pm $ and $s_{2}=\pm$, respectively. We have
\begin{equation}
\Theta _{1}^{s_{1},s_{2}}\left( \theta _{1}\right) =i_{\ell _{1}}\cos^{e_{1}}\theta _{1}\sin ^{e_{2}}\theta _{1}\mathbf{P}_{\ell _{1}-\left(e_{1}+e_{2}\right) /2}^{\left( \mu _{2}+e_{2}-1/2;\mu _{1}+e_{1}-1/2\right)
}\left( \cos 2\theta _{1}\right) ,  \label{sol1}
\end{equation}
where $i_{\ell _{1}}$ is a normalization constant, $\mathbf{P}_{n}^{\left(\alpha ;\beta \right) }\left( x\right) $ are Jacobi polynomials, and 
\begin{equation}
e_{j}=\left\{ 
\begin{array}{l}
1\text{ \ \ \ if }s_{j}=-1 \\ 
0\text{ \ \ \ if }s_{j}=+1
\end{array}%
\right. .
\end{equation}
The solution of Eq.~ (\ref{sol1}) corresponds to the eigenvalue $\lambda_{1}^{2}=4\ell _{1}\left( \ell _{1}+\mu _{1}+\mu _{2}\right) $.  Notably:
\begin{itemize}
\item When $s_{1}\cdot s_{2}=-1$, then $\ell _{1}$ takes on positive half-integer values (such as 1/2, 3/2, $\cdots$).
\item In contrast, when both $s_{1}=s_{2}=1$, then $\ell _{1}$ becomes a non-negative integer.
\item A special case for $\ell _{1}=0$,  where only the state with $s_{1}=s_{2}=1$ exists.
\end{itemize}
The angular solutions $\Theta _{2}\left( \theta _{2}\right) $ are characterized by the eigenvalue $s_{3}=\pm $ of the reflection operator $R_{3}$. One has:
\begin{equation}
\Theta _{2}^{s_{3}}\left( \theta _{2}\right) =i_{\ell _{2}}\cos
^{e_{3}}\theta _{1}\sin ^{2\ell _{1}}\theta _{2}\mathbf{P}_{\ell _{2}-\frac{%
e_{3}}{2}}^{\left( 2\ell _{1}+\mu _{1}+\mu _{2};\mu _{3}+e-1/2\right)
}\left( \cos 2\theta _{2}\right) .
\end{equation}
When $s_{3}=1$, $\ell _{2}$ takes on non-negative integer values, and in contrast for $s_{3}=-1$, $\ell _{2}$ assumes positive half-integer values. The separation constant $\lambda _{2}$ is given by
\begin{equation}
\lambda _{2}^{2}=4\left( \ell _{2}+\ell _{1}\right) \left( \ell _{2}+\ell
_{1}+\mu _{1}+\mu _{2}+\mu _{3}+1/2\right) .
\end{equation}%
In Table \ref{table:1}, we provide expressions for the angular components $\Theta _{j}^{s_{j+1}}\left( \theta _{j}\right) $
\begin{table}[H]
\centering
\begin{tabular}{||l|l||}
\hline
$j$ & $\Theta _{j}^{s_{j+1}}\left( \theta _{j}\right) $ \\ \hline
$3$ & $\cos ^{e_{4}}\theta _{3}\sin ^{2\left( \ell _{2}+\ell _{1}\right)
}\theta _{3}\mathbf{P}_{\ell _{3}-\frac{e_{4}}{2}}^{\left( 1/2+2\left( \ell
_{2}+\ell _{1}\right) +\mu _{1}+\mu _{2}+\mu _{3},\mu _{4}+e_{4}-1/2\right)
}\left( \cos 2\theta _{3}\right) $ \\ \hline
$4$ & $\cos ^{e_{5}}\theta _{2}\sin ^{2\left( \ell _{2}+\ell _{1}+\ell
_{3}\right) }\theta _{2}\mathbf{P}_{\ell _{4}-\frac{e_{5}}{2}}^{\left(
1+2\left( \ell _{1}+\ell _{2}+\ell _{3}\right) +\mu _{1}+\cdots +\mu
_{4},\mu _{5}+e_{5}-1/2\right) }\left( \cos 2\theta _{4}\right) $ \\ \hline
$5$ & $\cos ^{e_{6}}\theta _{5}\sin ^{2\left( \ell _{1}+\cdots +\ell
_{4}\right) }\theta _{5}\mathbf{P}_{\ell _{5}-\frac{e_{6}}{2}}^{\left(
3/2+2\left( \ell _{1}+\cdots +\ell _{4}\right) +\mu _{1}+\cdots +\mu
_{5},\mu _{6}+e_{6}-1/2\right) }\left( \cos 2\theta _{5}\right) $ \\ \hline
$6$ & $\cos ^{e_{7}}\theta _{6}\sin ^{2\left( \ell _{1}+\cdots +\ell
_{5}\right) }\theta _{6}\mathbf{P}_{\ell _{6}-\frac{e_{7}}{2}}^{\left(
2+2\left( \ell _{1}+..+\ell _{5}\right) +\mu _{1}+\cdots +\mu _{5},\mu
_{7}+e_{7}-1/2\right) }\left( \cos 2\theta _{6}\right) $ \\ \hline
$\vdots $ & $\vdots $ \\ \hline
$k$ & $\cos ^{e_{k+1}}\theta _{k}\sin ^{2\left( \ell _{1}+\cdots +\ell
_{k-1}\right) }\theta _{k}\mathbf{P}_{\ell _{k}-\frac{e_{k+1}}{2}}^{\left( 
\frac{k-2}{2}+2\left( \ell _{1}+..+\ell _{k-1}\right) +\mu _{1}+\cdots +\mu
_{k},\mu _{k+1}+e_{k+1}-1/2\right) }\left( \cos 2\theta _{k}\right) $ \\ 
\hline
\end{tabular}%
\caption{First few angular solutions $\Theta _{j}^{s_{j+1}}\left( \protect%
\theta _{j}\right) $.}
\label{table:1}
\end{table}
while in Table \ref{table:2}, we tabulate the separation constants
\begin{table}[H]
\centering
\begin{tabular}{||l|l||}
\hline
$j$ & $\lambda _{j}^{2}$ \\ \hline
$3$ & $4\left( \ell _{1}+\ell _{2}+\ell _{3}\right) \left( \ell _{1}+\ell
_{2}+\ell _{3}+\mu _{1}+\cdots +\mu _{4}+1\right) $ \\ \hline
$4$ & $4\left( \ell _{1}+\ell _{2}+\ell _{3}+\ell _{4}\right) \left( \ell
_{1}+\cdots +\ell _{4}+\mu _{1}+\cdots +\mu _{5}+3/2\right) $ \\ \hline
$5$ & $4\left( \ell _{1}+\cdots +\ell _{5}\right) \left( \ell _{1}+\cdots
+\ell _{5}+\mu _{1}+\cdots +\mu _{6}+2\right) $ \\ \hline
$6$ & $4\left( \ell _{1}+\ell _{2}+\cdots +\ell _{6}\right) \left( \ell
_{1}+\ell _{2}+\cdots +\ell _{6}+\mu _{1}+\cdots +\mu _{7}+5/2\right) $ \\ 
\hline
$\vdots $ & $\vdots $ \\ \hline
$k$ & $4\left( \ell _{1}+\ell _{2}+\cdots +\ell _{k}\right) \left( \ell
_{1}+\ell _{2}+\cdots +\ell _{k}+\mu _{1}+\cdots +\mu _{k+1}+\frac{k-1}{2}%
\right) $ \\ \hline
\end{tabular}%
\caption{First few separation constant $\protect\lambda _{j}^{2}$.}
\label{table:2}
\end{table}
Next, we have to seek an exact solution to the radial equation:
\begin{equation}
\left[ \frac{\partial ^{2}}{\partial r^{2}}+\frac{d-1+2\left( \mu _{1}+\mu_{2}+\mu _{3}+\cdots+\mu _{d}\right) }{r}\frac{\partial }{\partial r}+E^{2}-m^{2}-\frac{\varpi ^{2}}{r^{2}}\right] \mathcal{R}\left( r\right) =0,
\label{rad}
\end{equation}
where the separation constant $\varpi ^{2}$ is 
\begin{equation}
\varpi ^{2}=4\left( \ell _{1}+\ell _{2}+\cdots +\ell _{d-1}\right) \left(
\ell _{1}+\ell _{2}+\cdots +\ell _{d-1}+\mu _{1}+\cdots +\mu _{d}+\frac{d-2}{%
2}\right) .
\end{equation}%
In the following section, we apply Eq.~ (\ref{rad}) to analyze the DKG equation in two key scenarios: the harmonic oscillator and a Coulomb-like potential. Our primary goal is to derive the energy eigenvalues and corresponding eigenfunctions for both cases, providing deeper insights into the quantum mechanical behavior influenced by Dunkl operators and their impact on the dynamics of systems governed by these potentials.

\section{Dunkl-Klein-Gordon oscillator}
\label{sec3}
For stationary states, the Dunkl-oscillator in the Klein-Gordon equation in d-dimensional Cartesian coordinates can be expressed as
\begin{equation}
\left[ E^{2}-\left( \frac{1}{i}D_{j}+im\omega x_{j}\right) \left( \frac{1}{i}%
D_{j}-im\omega x_{j}\right) -m^{2}\right] \Psi \left( \mathbf{x}\right) =0,
\end{equation}%
where $j=1,2,\cdots,d$. This equation can also be written in the form:
\begin{equation}
\Bigg[\Big(D_{1}^{2}+\cdots +D_{d}^{2}\Big)+2m\omega \bigg( \mu _{1}R_{1}+\cdots +\mu_{d}R_{d}+\frac{d}{2}\bigg) -m^{2}\omega ^{2}\Big( x_{1}^{2}+\cdots +x_{d}^{2}\Big) +E^{2}-m^{2}\Bigg]\Psi \left( \mathbf{x}\right) =0, \label{dKGO}
\end{equation}
where $m$ and $\omega $ represent the rest mass and oscillator frequency, respectively. We observe that Eq.~ \eqref{dKGO} differs from Eq.~ \eqref{kg01} due to the two terms within the second and third parentheses.  Consequently, the radial component of the DKG oscillator in d-dimensions can be expressed as
\begin{eqnarray}
\bigg[ \frac{d^{2}}{dr^{2}}+\frac{d-1+2\left( \mu _{1}+\mu _{2}+\mu_{3}+\cdots+\mu _{d}\right) }{r}\frac{d}{dr}&+&2m\omega \left( \mu_{1}s_{1}+\cdots +\mu _{d}s_{d}+\frac{d}{2}\right) \nonumber  \\
&-&m^{2}\omega^{2}r^{2}+E^{2}-m^{2}-\frac{\varpi ^{2}}{r^{2}}\bigg] \mathcal{R}\left(r\right) =0.  \label{an}
\end{eqnarray}
To derive the solution of the radial part, we first suggest a radial coordinate transformation, 
\begin{equation}
\rho =m\omega r^{2}.
\end{equation}
Substituting this into Eq.~ (\ref{an}), we obtain
\small
\begin{eqnarray}
\left[ \rho \frac{d^{2}}{d\rho ^{2}}+\left( \frac{d}{2}+\mu _{1}+\mu
_{2}+\cdots+\mu _{d}\right) \frac{d}{d\rho }-\frac{\rho }{4}-\frac{%
\varpi ^{2}}{4\rho }+\frac{1}{2}\left( \mu _{1}s_{1}+\cdots +\mu _{d}s_{d}+%
\frac{d}{2}\right) +\frac{E^{2}-m^{2}}{4m\omega }\right] \mathcal{R}\left(
\rho \right) =0. \label{1n}
\end{eqnarray}
\normalsize
To analyze the asymptotic behavior of the wave function at both the origin
and infinity, and to determine the existence of regular solutions, we consider a solution of the following form:
\begin{equation}
\mathcal{R}\left( \rho \right) =e^{-\frac{\rho }{2}}\rho ^{\ell _{1}+\ell
_{2}+\cdots +\ell _{d-1}}\digamma \left( \rho \right) .
\end{equation}%
Substituting this form into Eq.~ (\ref{1n}) yields 
\begin{eqnarray}
&&\Bigg[ \rho \frac{d^{2}}{d\rho^{2}} 
+ \left(2\left( \ell _{1}+\ell _{2}+\cdots +\ell _{d-1}\right)+ \frac{d}{2} + \left(\mu_{1} + \mu_{2} + \cdots + \mu_{d} \right)-\rho\right) \frac{d}{d\rho} -\left( \ell _{1}+\ell _{2}+\cdots +\ell _{d-1}\right) \nonumber \\
&&+ \frac{E^{2} - m^{2}}{4m\omega}-\frac{1}{2} \Big(\left(\mu_{1} + \mu_{2} + \cdots + \mu_{d} \right)-\left(\mu_{1}s_1 + \mu_{2}s_2 + \cdots + \mu_{d}s_d \right)\Big)\Bigg] \digamma \left( \rho \right) =0.
\end{eqnarray}
This equation is identified as the confluent hypergeometric equation \cite{Gradshteyn}, and its solutions are expressed in terms of confluent hypergeometric functions. Consequently, the solution of Eq.~ \eqref{an} can be written as:
\begin{equation}
\mathcal{R}\left( \rho \right) =\mathcal{C}\text{ \ }\rho ^{\ell _{1}+\ell_{2}+\cdots +\ell _{d-1}}e^{-\frac{\rho }{2}}\mathbf{F}\left( a,b;\rho\right) ,  \label{pot3}
\end{equation}%
where $\mathcal{C}$ is the normalization constant and 
\begin{equation}
\left\{ 
\begin{array}{l}
a=\left(\ell _{1}+\ell _{2}+\cdots +\ell _{d-1}\right)+\frac{1}{2} \big[\left(\mu_{1} + \mu_{2} + \cdots + \mu_{d} \right)-\left(\mu_{1}s_1 + \mu_{2}s_2 + \cdots + \mu_{d}s_d \right)\big]-\frac{E^{2}-m^{2}%
}{4m\omega } \\ 
b=2\left(\ell _{1}+\ell _{2}+\cdots +\ell _{d-1}\right)+\frac{d}{2}+\left(\mu _{1}+\cdots+\mu
_{d}\right)
\end{array}%
\right. .
\end{equation}
It is crucial to emphasize that the solution of Eq.~ (\ref{pot3}) must be a polynomial of degree $n$. Achieving a finite polynomial solution is only feasible if the factor $a$ is a negative integer. Therefore we write
\begin{equation}
\ell _{1}+\ell _{2}+\cdots +\ell _{d-1}+\frac{\mu _{1}+\mu _{2}+\cdots+\mu
_{d}-\left( \mu _{1}s_{1}+\cdots +\mu _{d}s_{d}\right) }{2}-\frac{E^{2}-m^{2}%
}{4m\omega }=-n.
\end{equation}%
Using this condition, we obtain the quantized energy spectrum of the DKG oscillator as follows:
\begin{equation}
E_{n,s_{1},\cdots,s_{d}}=\pm \sqrt{2m\omega \left[ 2\left( n+\ell _{1}+\ell
_{2}+\cdots +\ell _{d-1}\right) +\mu _{1}+\mu _{2}+\cdots+\mu _{d}-\left( \mu
_{1}s_{1}+\cdots +\mu _{d}s_{d}\right) \right] +m^{2}}.\label{EDO}
\end{equation}%
Let us examine the result presented in Eq.~ (\ref{EDO}). Firstly, it is important to point out that for $d=3$, the derived energy spectrum reduces to that of the DKG oscillator in three spatial dimensions, consistent with previous expectations  \cite{Bilel20221}. Secondly, it is observed that the energy explicitly depends on the deformation constant $\mu$ and the eigenvalues of the reflection operator. Specifically, due to the inclusion of the Dunkl derivative, the energy spectrum is influenced by the reflection operator, indicating that the spectrum is sensitive to parity.

\begin{figure}[htb!]
\begin{minipage}[t]{0.5\textwidth}
        \centering
        \includegraphics[width=\textwidth]{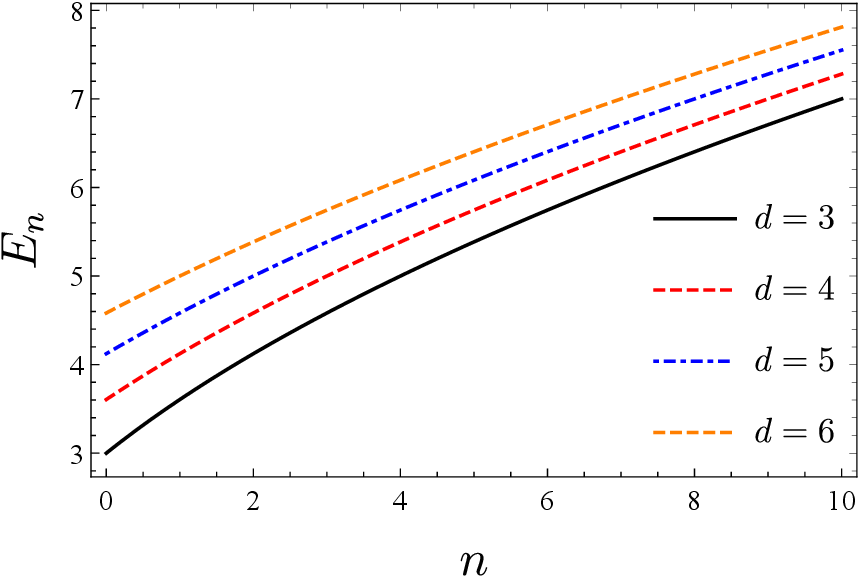}
       \subcaption{ $s_{i}=+1 $}\label{figE1}
   \end{minipage}%
\begin{minipage}[t]{0.50\textwidth}
        \centering
       \includegraphics[width=\textwidth]{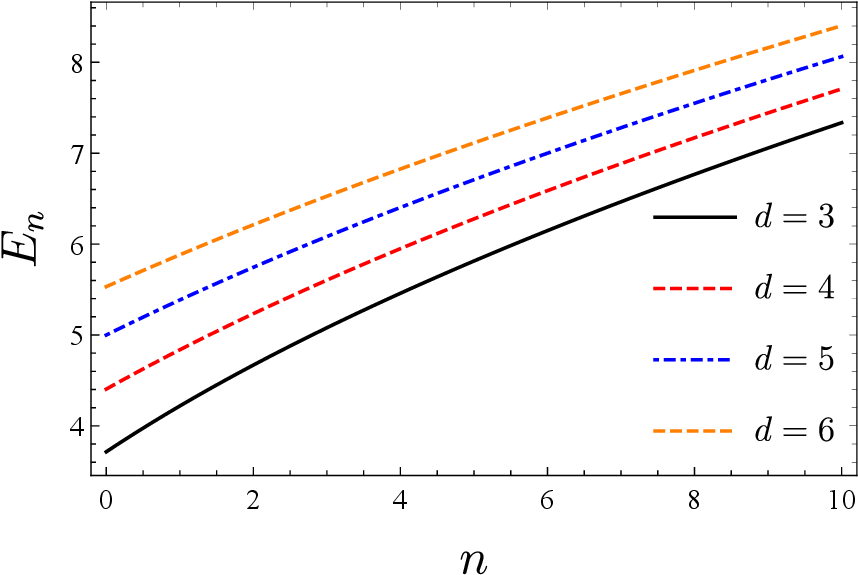}\\
        \subcaption{$s_{i}=-1 $}\label{fig:E2}
    \end{minipage}\hfill
\caption{DKG-oscillator spectra versus  $n$, where $\ell_{i}=1$, $\mu_{i}=0.4$. }
\label{Efigsnew}
\end{figure}

\begin{figure}[htb!]
\begin{minipage}[t]{0.5\textwidth}
        \centering
        \includegraphics[width=\textwidth]{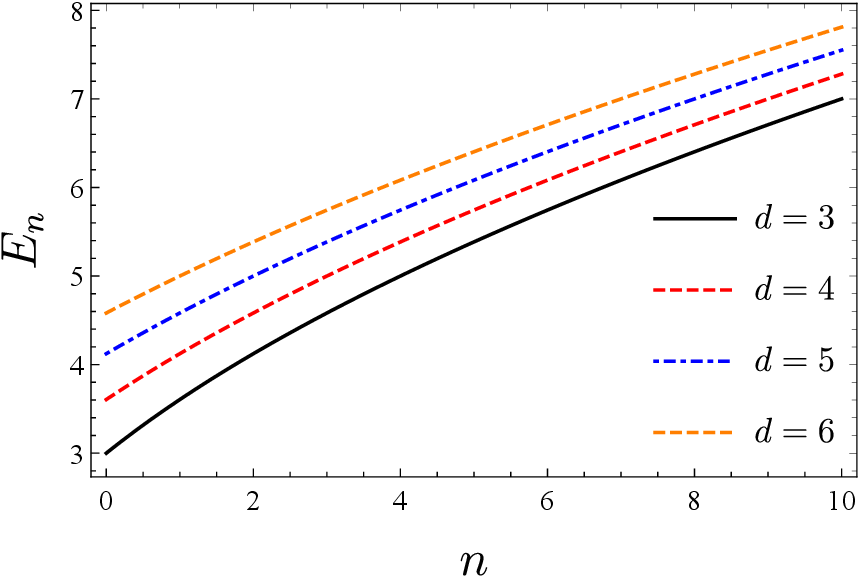}
       \subcaption{ $s_{i}=+1 $}\label{figEC3}
   \end{minipage}%
\begin{minipage}[t]{0.50\textwidth}
        \centering
       \includegraphics[width=\textwidth]{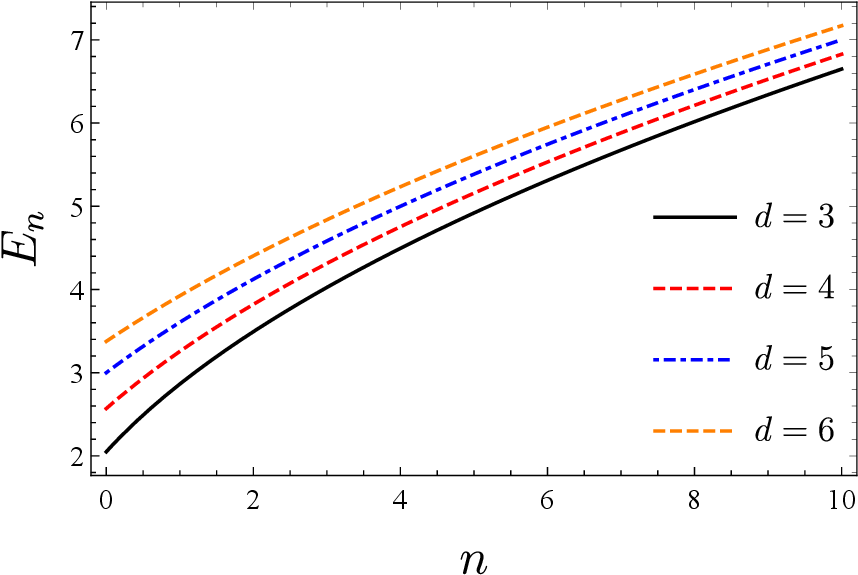}\\
        \subcaption{$s_{i}=-1 $}\label{fig:E4}
    \end{minipage}\hfill
\caption{DKG-oscillator spectra versus  $n$, where $\ell_{i}=1$, $\mu_{i}=-0.4$. }
\label{Efigs}
\end{figure}

In the non-relativistic limit, following the standard approach of setting $E=m+E_{nr}$ and assuming $m \gg E_{nr}$, one can perform a Taylor expansion of Eq.~ (\ref{EDO}). It yields
\begin{equation}
E_{nr}=\omega \Big[ 2\big( n+\ell _{1}+\ell _{2}+\cdots +\ell_{d-1}\big) +\big(\mu _{1}+\mu _{2}+\cdots+\mu _{d}\big)-\big( \mu _{1}s_{1}+\cdots
+\mu _{d}s_{d}\big) \Big] .
\end{equation}
To explore the influence of spatial dimensions and parity parameters on the energy levels and system dynamics, we generate Fig. \ref{Efigsnew} and Fig. \ref{Efigs} by plotting the energy levels as a function of quantum numbers for both positive and negative parity with positive and negative Dunkl parameter value, respectively. In this analysis, we use the sub-figures, e.g. Fig. \ref{figE1} and Fig. \ref{figEC3}, to illustrate the effect of the Dunkl parameter when the parity is positive, and similarly Fig. \ref{fig:E2} and Fig. \ref{fig:E4} to display the same effect when the parity is negative. It is worth mentioning that throughout the figures in this manuscript, the two key parameters, mass and oscillator frequency, are consistently set to unity.

\newpage

The analysis of these figures reveals the following conclusions:
\begin{itemize}

\item The energy levels exhibit a monotonic increase with respect to the quantum number $n$. Additionally, for a fixed value of $n$, the energy levels rise as the spatial dimension $d$ increases.


\item When $\mu_{i} = +0.4$, the energy associated with even parity states is lower than that of odd parity states. Conversely, for $\mu_{i} = -0.4$, the energy of the even parity states is greater than that of the odd parity states.


\end{itemize}

\begin{figure}[H]
\begin{minipage}[t]{0.5\textwidth}
        \centering
        \includegraphics[width=\textwidth]{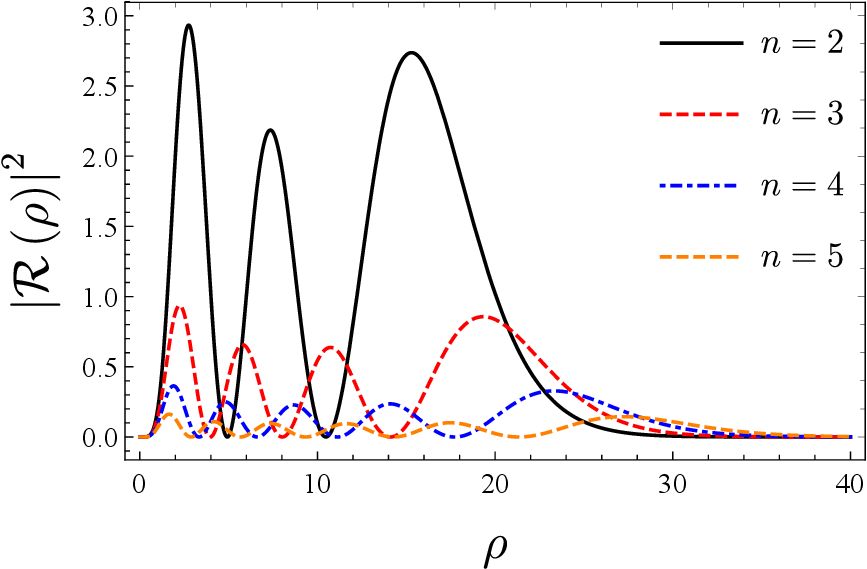}
       \subcaption{ $\mu_{i}=+0.4 $}\label{fig:fun1}
   \end{minipage}%
\begin{minipage}[t]{0.50\textwidth}
        \centering
       \includegraphics[width=\textwidth]{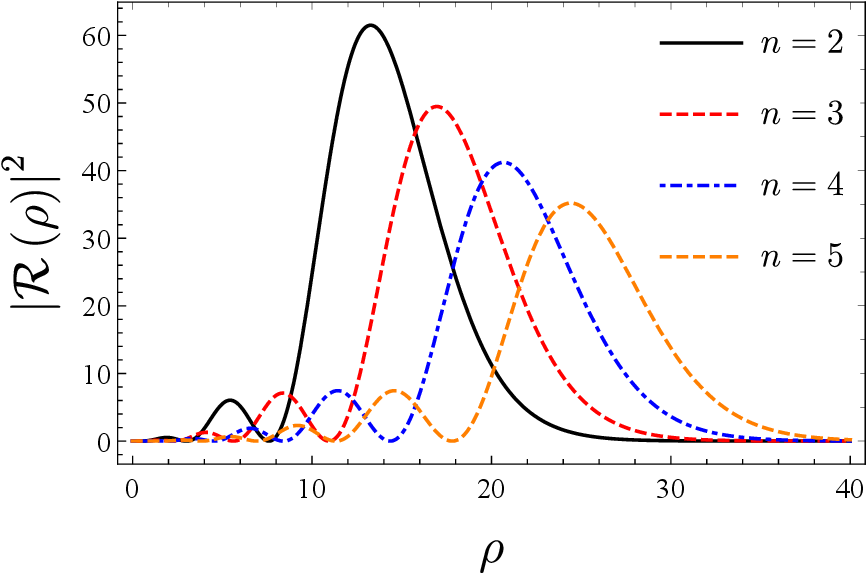}\\
        \subcaption{$\mu_{i}=-0.4 $}\label{fig:fun2}
    \end{minipage}\hfill
\caption{The square modulus of the radial solution as a function of $\rho$ for four levels $n = 2, 3,4,5$.}
\label{Efunc1}
\end{figure}

Then, in Fig. \ref{Efunc1}, we depict the square modulus of the radial solution as a function of the radial distance $\rho$ for several quantum levels for both positive and negative Dunkl parameter values, respectively. 

\newpage

The behavior of $\left\vert \mathcal{R}\left( \rho \right) \right\vert ^{2}$ is clearly influenced by both the quantum number and the Dunkl parameter. Our observations are as follows:
\begin{itemize}

\item For a fixed quantum number $n$, changes in the Dunkl parameter $\mu_{i}$ significantly affect the intensity of the peaks. 

\item For a constant Dunkl parameter $\mu_{i}$, the intensity of the peaks decreases as the quantum number $n$ increases. 

\item All peaks are symmetric around a specific radial position $\rho$.

\end{itemize}

\section{Coulomb potential}
\label{sec4}

In this section, we address the second problem by examining the solutions for bound and scattering states of the d-dimensional DKG equation with a Dunkl-Coulomb-like potential.

\subsection{Bound state}

We begin with an analysis of the bound state solutions within the Dunkl formalism, which is essential for understanding a range of physical phenomena, from atomic structure to elementary particle physics. To solve the equation, we incorporate an attractive Coulomb-type potential
\begin{eqnarray}
 V=\frac{-Ze^{2}}{r}.  
\end{eqnarray}
We begin by incorporating this potential into the equation of motion through substitution into Eq.~ (\ref{rad}). The resulting radial equation reads:
\begin{equation}
\left[ \frac{d^{2}}{dr^{2}}+\frac{d-1+2\left( \mu _{1}+\mu _{2}+\mu_{3}+\cdots+\mu _{d}\right) }{r}\frac{d}{dr}+\left( E+\frac{Ze^{2}}{r}\right)^{2}-m^{2}-\frac{\varpi ^{2}}{r^{2}}\right] \mathcal{R}\left( r\right) =0. \label{34}
\end{equation}
To solve this equation, we assume the following ansazt:
\begin{equation}
\mathcal{R}=\eta ^{\delta }e^{-\frac{\eta }{2}}\Xi \left( \eta \right) ,
\end{equation}%
where 
\begin{equation}
\eta =2\varkappa r \qquad {\text{and}} \qquad \varkappa =\sqrt{m^{2}-E^{2}},
\end{equation}
with
\begin{eqnarray}
\delta =1-\allowbreak \frac{d+2\left( \mu _{1}+\cdots+\mu _{d}\right) }{2}+\sqrt{\varpi ^{2}+\left( \mu _{1}+\cdots+\mu _{d}\right) \left( \mu_{1}+\cdots+\mu _{d}+d-2\right) +\left( \frac{d}{2}-1\right) ^{2}-Z^{2}e^{4}}.
\end{eqnarray}%
Through a straightforward calculation, we find that $\Xi \left( \eta \right)$ satisfies
\small
\begin{eqnarray}
\Bigg[ \eta \frac{d^{2}}{d\eta ^{2}}+\Big( 2\delta +d-1+2\left( \mu_{1}+\cdots+\mu _{d}\right) -\eta \Big) \frac{d}{d\eta }+\frac{EZe^{2}}{\varkappa }-\frac{d-1+2\left( \mu _{1}+\mu _{2}+\mu _{3}+\cdots+\mu _{d}\right) }{2}-\delta \Bigg] \Xi \left( \eta \right) =0.\label{DCO}
\end{eqnarray}
\normalsize
This corresponds to the confluent hypergeometric equation. Based on the previous section results, we can express the eigenfunctions and the energy eigenvalues of Eq.~ (\ref{DCO}) as follows:
\begin{equation}
\Xi \left( \eta \right) =\mathbf{F}\Big( -n,2\delta +d-1+2\left( \mu_{1}+\cdots+\mu_{d}\right) ;\eta \Big) ,
\end{equation}%
\begin{equation}
\frac{E_{n}}{m}=\left\{ 1+\frac{Z^{2}e^{4}}{\left( n-\allowbreak \frac{1}{2}-\sqrt{\varpi ^{2}+\left( \mu _{1}+\cdots+\mu _{d}\right) \left( \mu_{1}+\cdots+\mu _{d}+d-2\right) +\left( \frac{d}{2}-1\right) ^{2}-Z^{2}e^{4}}%
\right) ^{2}}\right\} ^{-1/2}. \label{coe}
\end{equation}
Now, let us briefly express the implications of Eq.~ (\ref{coe}):

\begin{itemize}

\item For the case where $d=3$,  Eq.~ \eqref{coe} reduces to the energy spectrum of the Dunkl-Coulomb potential in three spatial dimensions, as anticipated.

\item We observe that for the Coulomb-type radial potential,  the energy spectrum is independent of the eigenvalues of the reflection operator, meaning that the spectrum is not affected by parity.

\item A constrain in the following form
\begin{equation}
\varpi ^{2}+\left( \mu _{1}+\cdots+\mu _{d}\right) \left( \mu _{1}+\cdots+\mu
_{d}+d-2\right) +\left( \frac{d}{2}-1\right) ^{2}-Z^{2}e^{4}\geq 0,
\end{equation}
is necessary for ensuring the existence of physical energy eigenvalues.

\end{itemize}

\begin{figure}[htb]
\begin{minipage}[t]{0.5\textwidth}
        \centering
        \includegraphics[width=\textwidth]{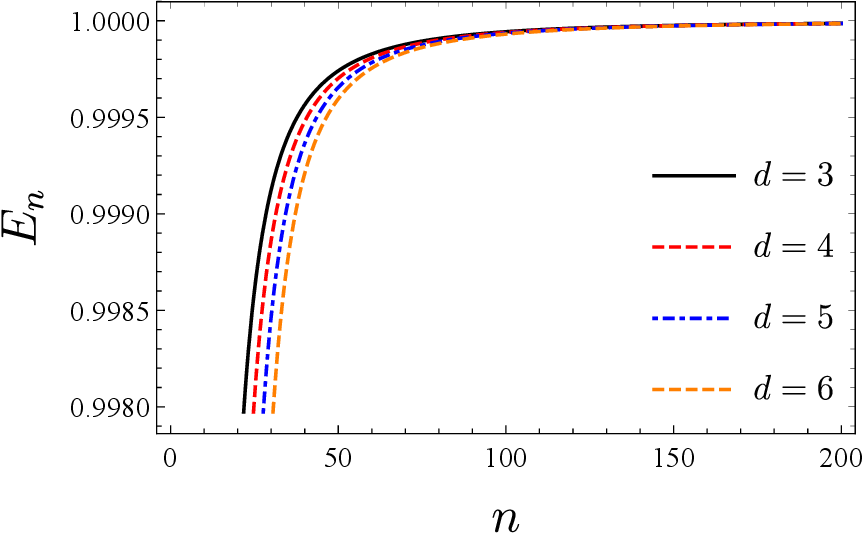}
       \subcaption{ $\mu_{i}=0.4 $}\label{fig:Ec1}
   \end{minipage}%
\begin{minipage}[t]{0.50\textwidth}
        \centering
       \includegraphics[width=\textwidth]{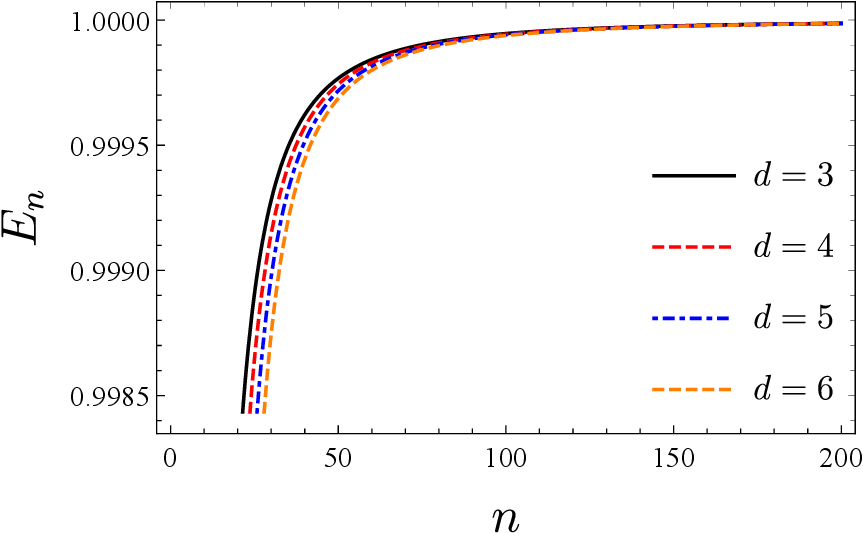}\\
        \subcaption{$\mu_{i}=-0.4 $}\label{fig:Ec2}
    \end{minipage}\hfill
\caption{The energy levels of the Dunkl Coulomb potential vs. $n$ in different dimensions, where $\ell_{i}=Ze^{2}=1$.}
\label{Efigc1}
\end{figure}

We now proceed with a graphical analysis to illustrate the effect of the Dunkl parameter.  In Fig. \ref{Efigc1}, we present the energy levels of the Dunkl Coulomb potential as a function of the quantum number $n$ for various spatial dimensions $d$. The plot reveals that the energy levels initially increase rapidly with $n$, but the rate of increase diminishes for larger values of $n$. Additionally, for a fixed $n$, the energy levels decrease as the spatial dimension $d$ increases.

\begin{figure}[htb]
\begin{minipage}[t]{0.5\textwidth}
        \centering
        \includegraphics[width=\textwidth]{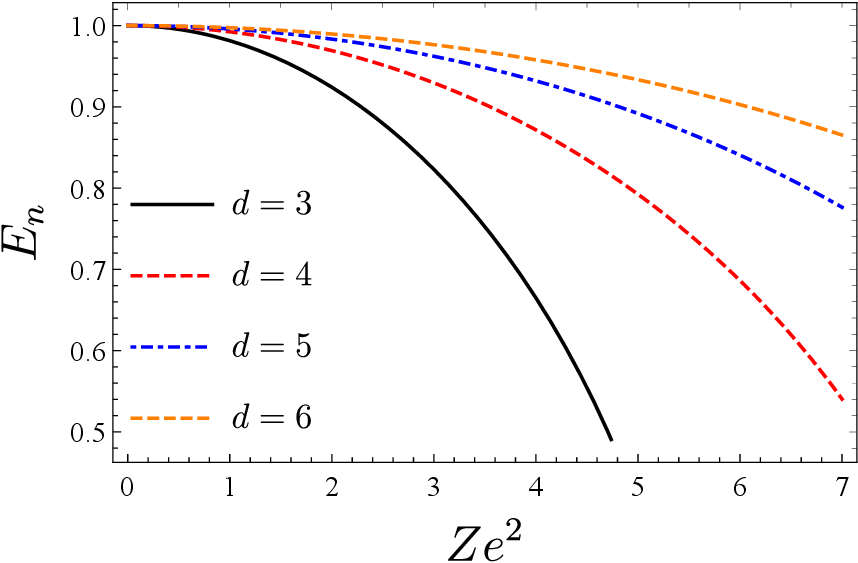}
       \subcaption{ $\mu_{i}=0.4$}\label{fig:Ec3}
   \end{minipage}%
\begin{minipage}[t]{0.50\textwidth}
        \centering
       \includegraphics[width=\textwidth]{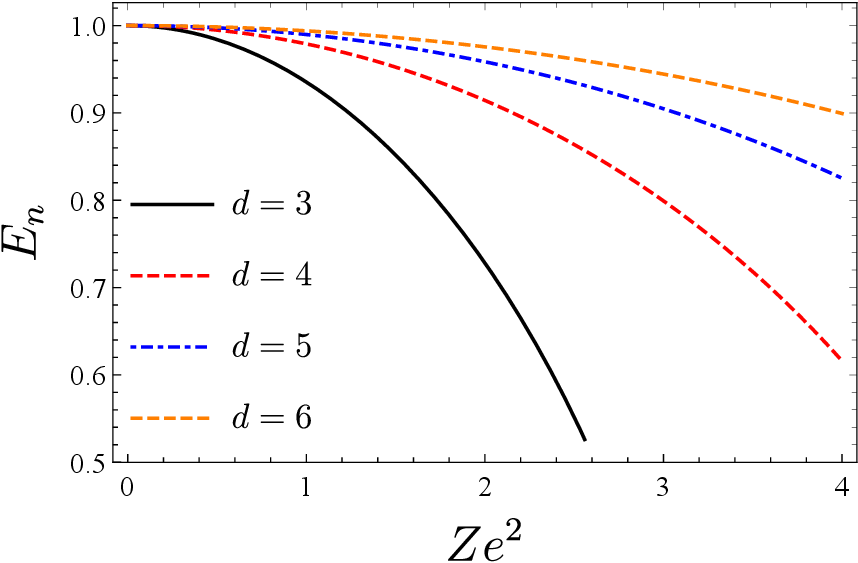}\\
        \subcaption{$\mu_{i}=-0.4$}\label{fig:Ec4}
    \end{minipage}\hfill
\caption{The energy levels of the Dunkl Coulomb potential vs. $Ze^{2}$ in different dimensions, where $\ell_{i}=n=1$.}
\label{Efigc2}
\end{figure}

Next, in Fig. \ref{Efigc2}, we plot the energy levels with respect to $Ze^{2}$. For these numerical values, there is an accumulation point at $Z_{cr}$, where all curves tend to the minimum energy value.

\begin{figure}[H]
\begin{minipage}[t]{0.5\textwidth}
        \centering
        \includegraphics[width=\textwidth]{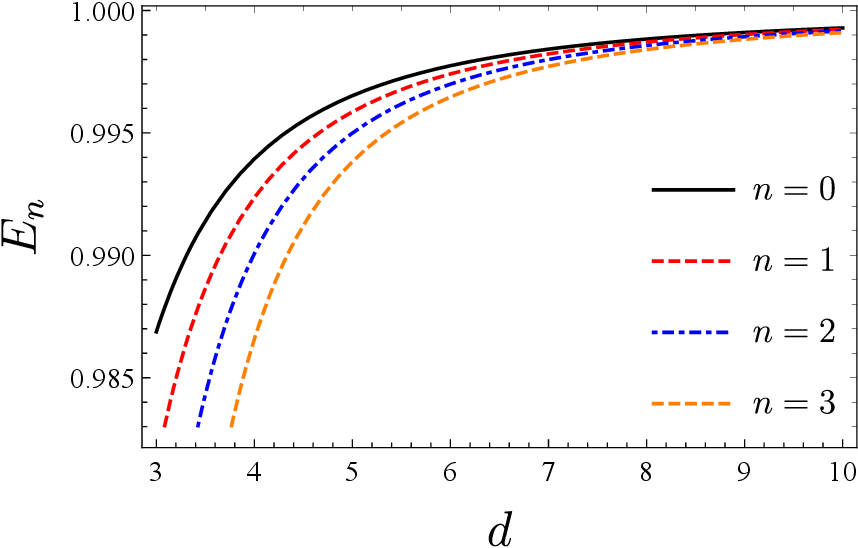}
       \subcaption{ $\mu_{i}=0.4$}\label{fig:Ec5}
   \end{minipage}%
\begin{minipage}[t]{0.50\textwidth}
        \centering
       \includegraphics[width=\textwidth]{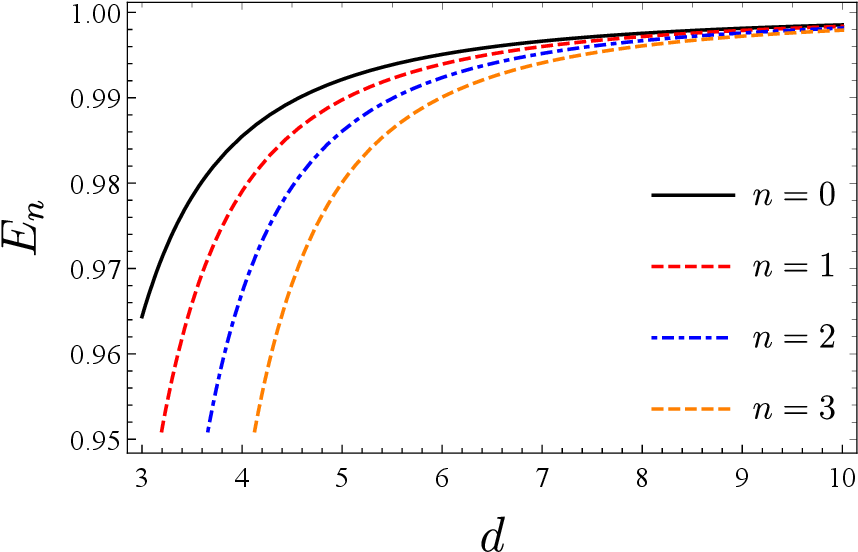}\\
        \subcaption{$\mu_{i}=-0.4$}\label{fig:Ec6}
    \end{minipage}\hfill
\caption{The energy levels of the Dunkl Coulomb potential as a function of spatial dimension for different values of $n$, where $\ell_{i}=Ze^{2}=1$.}
\label{Efigc3}
\end{figure}

Finally, in Fig. \ref{Efigc3}, we present the energy levels of the Dunkl Coulomb potential as a function of the spatial dimension for four quantum states, $n = 0, 1, 2, 3$. This figure demonstrates the influence of spatial dimensionality on the energy levels.


\subsection{Scattering states}
External field problems constitute a distinct class of challenges in quantum field theory, prominently characterized by the phenomenon of particle creation—one of the most striking aspects of quantum fields \cite{Sauter, Heisenberg, Schwinger}. Physically, particle creation occurs due to the interaction between external fields and the vacuum state vector, causing its modification and leading to particle emission. Several methods are available to compute the pair creation rate, including the Feynman path-integral approach \cite{Hawking, Chitre}, the Schwinger method \cite{Schwinger}, the Hamiltonian diagonalization technique \cite{Grib}, and the "in" and "out" formalism \cite{Nikishov}, which is employed in this subsection. To address this problem, we begin by proposing a radial coordinate transformation, \begin{equation} 
\zeta =-2i\kappa r, \label{eq45}
\end{equation} 
where $\kappa =\sqrt{E^{2}-m^{2}}$. By substituting Eq.~ \eqref{eq45} into Eq.~ (\ref{34}), we obtain
\begin{equation}
\left[ \frac{\partial ^{2}}{\partial \zeta ^{2}}+\frac{d-1+2\left( \mu
_{1}+\mu _{2}+\mu _{3}+\cdots+\mu _{d}\right) }{\zeta }\frac{\partial }{%
\partial \zeta }-\frac{i}{\kappa }\frac{EZe^{2}}{\zeta }+\frac{%
Z^{2}e^{4}-\varpi ^{2}}{\zeta ^{2}}-\frac{1}{4}\right] \mathcal{R}\left(
r\right) =0.  \label{Su1}
\end{equation}%
Then, to reduce this equation to a well-known class of differential equations, we apply the following transformation%
\begin{equation}
\mathcal{R}\left( \zeta \right) =\zeta ^{\vartheta }\Phi \left( \zeta
\right) .  \label{Su}
\end{equation}%
Substituting Eq.~ (\ref{Su}) into (\ref{Su1}), we arrive at:
\begin{equation}
\left[ \frac{d^{2}}{d\zeta ^{2}}-\frac{i}{\kappa }\frac{EZe^{2}}{\zeta }-%
\frac{1}{4}+\frac{\left( Ze^{2}\right) ^{2}-\vartheta \left( \vartheta
+1\right) -\varpi ^{2}}{\zeta ^{2}}\right] \Phi \left( \zeta \right) =0,
\label{eqs1}
\end{equation}
where
\begin{equation}
2\vartheta =-\left[ d-1+2\left( \mu _{1}+\cdots+\mu _{d}\right) \right] .
\end{equation}
Eq.~ (\ref{eqs1}) is a second-order differential equation, which implies the existence of two independent solutions. Any other solution can be expressed as a linear combination of these two fundamental solutions. Now, our objective is to identify two sets of independent solutions, where :

\begin{description}

\item[-] The first set,  including $\Phi _{in}^{+}\left( \zeta \right) $ and $\Phi _{in}^{-}\left( \zeta \right) $, should behave like positive and
negative energy states as $\zeta \rightarrow 0$.

\item[-] The second set, consisting of \ $\Phi _{out}^{+}\left( \zeta \right) $and $\Phi _{out}^{-}\left( \zeta \right) $, should resemble positive
and negative energy states as $\zeta \rightarrow +\infty $.
\end{description}

Before deriving the exact solutions to Eq.~ (\ref{eqs1}), it is essential to analyze its behavior in two limiting cases: as $\zeta \rightarrow 0$ and as $\zeta \rightarrow +\infty $. These asymptotic behaviors can be straightforwardly determined via:
\begin{enumerate}
    \item For $\zeta \rightarrow +\infty $, Eq.~ (\ref{eqs1}) reduces to%

\begin{equation}
\left[ \frac{d^{2}}{d\zeta ^{2}}-\frac{1}{4}\right] \Phi \left( \zeta\right) =0.
\end{equation}
Thus,
\begin{equation}
    \Phi \left( \zeta \right) \simeq e^{\frac{-\zeta }{2}}.
\end{equation}

    \item For $\zeta \rightarrow 0$, Eq.~ (\ref{eqs1}) becomes%
\begin{equation}
\left[ \frac{d^{2}}{d\zeta ^{2}}-\frac{\varpi ^{2}+\vartheta \left(\vartheta +1\right) -\left( Ze^{2}\right) ^{2}}{\zeta ^{2}}\right] \Phi \left( \zeta \right) =0.
\end{equation}
So that, 
\begin{equation}
\Phi \left( \zeta \right) \simeq \zeta ^{\frac{1}{2}\pm i\sqrt{\left( Ze^{2}\right) ^{2}-\varpi^{2}-\vartheta \left( \vartheta +1\right) -\frac{1}{4}}}.
\end{equation}
\end{enumerate}
On the other hand, the exact solutions of the differential equation given by Eq.~ (\ref{eqs1}) can be expressed in terms of the Whittaker functions as:
\begin{equation}
\Phi \left( \zeta \right) =C_{1}W_{\alpha ,\beta }\left( \zeta \right)
+C_{2}M_{\alpha ,\beta }\left( \zeta \right) ,
\end{equation}
where $\alpha $ and $\beta $ are defined as:
\begin{equation}
\alpha =-i\frac{EZe^{2}}{\kappa };\qquad \beta =\pm i\sqrt{\left(
Ze^{2}\right) ^{2}-\varpi ^{2}-\vartheta \left( \vartheta +1\right) -\frac{1%
}{4}}.
\end{equation}
To determine the particle number density and the pair creation probability, we need to analyze the asymptotic behavior of the wave function $\Phi \left( \zeta \right) $ as $\zeta \rightarrow \infty $ and $\zeta\rightarrow 0$. This involves applying the Bogoliubov transformation technique \cite{Nikishov}, which connects the positive frequency solution $\Phi _{in}^{+}$ as $\zeta \rightarrow 0$, with the positive frequency solution $\Phi_{out}^{+}$ and its complex conjugate $\Phi _{out}^{-}$ as $\zeta
\rightarrow \infty $, as follows:
\begin{equation}
\Phi _{in}^{+}=A\Phi _{out}^{+}+B\Phi _{out}^{-},
\end{equation}%
where $A$ and $B$ are the Bogoliubov coefficients, which encode information about particle pair creation and satisfy the following relation
\begin{equation}
\left\vert A\right\vert ^{2}-\left\vert B\right\vert ^{2}=1.  \label{cond}
\end{equation}%
Now, let us examine the asymptotic behavior of the $W_{\alpha ,\beta }\left(z\right) $ as $\zeta \rightarrow \infty ,$
\begin{equation}
\lim_{\zeta \rightarrow \infty }W_{\alpha ,\beta }\left( \zeta \right)
\simeq \zeta ^{\alpha }e^{\frac{-\zeta }{2}}.
\end{equation}%
Consequently, we obtain that the negative and positive frequency solutions as $\zeta\rightarrow \infty $ reads:
\begin{equation}
\Phi _{out}^{-}\left( \zeta \right) =W_{\alpha ,\beta }\left( \zeta \right) ,
\end{equation}%
\begin{equation}
\Phi _{out}^{+}\left( \zeta \right) =\left( W_{\alpha ,\beta }\left( \zeta
\right) \right) ^{\ast }=W_{-\alpha ,\beta }\left( -\zeta \right) .
\end{equation}%
Analogously, by examining the behavior of $M_{\alpha ,\beta }\left( \zeta
\right) $ as $\zeta \rightarrow 0,$%
\begin{equation}
\lim_{\zeta \rightarrow 0}M_{\alpha ,\beta }\left( \zeta \right) =\zeta
^{1/2+\beta }e^{\frac{-\zeta }{2}},
\end{equation}%
we find that the corresponding positive frequency modes as $\zeta
\rightarrow 0$:
\begin{equation}
\Phi _{in}^{+}\left( \zeta \right) =M_{\alpha ,\beta }\left( \zeta \right) .
\end{equation}%
By using the following property of Whittaker functions \cite{Gradshteyn}
\begin{equation}
M_{\alpha ,\beta }\left( \zeta \right) =\frac{\Gamma \left( 1+2\beta \right) 
}{\Gamma \left( 1/2+\beta -\alpha \right) }e^{i\pi \alpha }W_{-\alpha ,\beta
}\left( -\zeta \right) +\frac{\Gamma \left( 1+2\beta \right) e^{i\pi \left(
\alpha -\beta -1/2\right) }}{\Gamma \left( 1/2+\beta +\alpha \right) }%
W_{\alpha ,\beta }\left( \zeta \right) ,
\end{equation}%
we express the positive frequency solution $\Phi _{in}^{+}\left( \zeta
\right) $ in terms of $\Phi _{out}^{-}\left( \zeta \right) $
and $\Phi _{out}^{+}\left( \zeta \right) $ as follows:
\begin{equation}
\Phi _{in}^{+}\left( \zeta \right) =\frac{\Gamma \left( 1+2\beta \right) }{\Gamma \left( 1/2+\beta -\alpha \right) }e^{i\pi \alpha }\Phi_{out}^{+}\left( \zeta \right) +\frac{\Gamma \left( 1+2\beta \right) e^{i\pi \left( \alpha -\beta -1/2\right) }}{\Gamma \left( 1/2+\beta +\alpha \right) } \Phi _{out}^{-}\left( \zeta \right) .
\end{equation}%
Then, using the Bogoliubov transformation that connects the 'in' and 'out' states along with the properties of Whittaker functions, we obtain the Bogoliubov coefficients in the form of
\begin{eqnarray}
A&=&\frac{\Gamma \left( 1+2\beta \right) }{\Gamma \left( 1/2+\beta -\alpha
\right) }e^{i\pi \alpha }, \\
B&=&\frac{\Gamma \left( 1+2\beta \right) e^{i\pi \left( \alpha -\beta
-1/2\right) }}{\Gamma \left( 1/2+\beta +\alpha \right) }.
\end{eqnarray}
In order to obtain the expressions of the probability we use the bosonic
condition given in Eq.~ (\ref{cond}) with the definition 
\begin{equation}
\mathcal{P}=\left\vert \frac{B}{A}\right\vert ^{2}.
\end{equation}%
Using the relation
\begin{equation}
\left\vert \Gamma \left( \frac{1}{2}+ix\right) \right\vert ^{2}=\frac{\pi }{%
\cosh \pi x},
\end{equation}%
the expression of the probability reads:
\begin{equation}
\mathcal{P}=\frac{\cosh \pi \left( \tilde{\beta}+\frac{EZe^{2}}{\kappa }%
\right) }{\cosh \pi \left( \tilde{\beta}-\frac{EZe^{2}}{\kappa }\right) }%
e^{-2\pi \tilde{\beta}}, \label{procon}
\end{equation}
where
\begin{equation}
\tilde{\beta}=\sqrt{\left( Ze^{2}\right) ^{2}-\varpi ^{2}-\vartheta \left(
\vartheta +1\right) -\frac{1}{4}}, \qquad\text{and}\quad \left( Ze^{2}\right)
^{2}-\varpi ^{2}-\vartheta \left( \vartheta +1\right) -\frac{1}{4}\geq 0%
\text{ }.\label{cond1}
\end{equation}
We now focus on the Dunkl-density of particles $\mathcal{N}$ created by Coulomb-like potential. For this quantity,  we use the following definition
\begin{equation}
\mathcal{N}=\Bigg( \left\vert \frac{B}{A}\right\vert ^{-2}-1\Bigg) ^{-1}.
\end{equation}%
After performing the calculations, we obtain the Dunkl-density of the created
particles 
\begin{equation}
\mathcal{N}=\frac{\cosh \pi \left( \tilde{\beta}+\frac{EZe^{2}}{\kappa }%
\right) }{e^{2\pi \tilde{\beta}}\cosh \pi \left( \tilde{\beta}-\frac{EZe^{2}%
}{\kappa }\right) -\cosh \pi \left( \tilde{\beta}+\frac{EZe^{2}}{\kappa }%
\right) }.
\end{equation}
Next, we examine the implications of the condition given in Eq.~ (\ref{cond1}) for the creation of spin-0 bosons, which is expressed as
\begin{equation}
Z^{2}e^{4}\geq \varpi ^{2}+\vartheta \left( \vartheta +1\right) +\frac{1}{4}.
\end{equation}%
In Dunkl-formalism in d-dimension this condition depends on several factors:
\begin{itemize}
    \item $Ze^{2}$,
    \item the quantum numbers $\ell _{j}$,
    \item the Dunkl parameters $\mu _{j}$,
    \item the spatial dimension $d$.
\end{itemize}
Additionally, we conclude that Eq.~ (\ref{procon}) can be utilized to establish a relationship between these factors as well as the most probable transitions when considering both small and large values of $Ze^2$, $\ell _{j}$, $\mu _{j}$, and $d$.
It is important to note that in the case where $\mu_{j} = 0$, corresponding to the absence of the Dunkl formalism, the aforementioned condition simplifies to:
\begin{equation}
Z\geq \allowbreak \frac{1}{e^{2}}\left( \ell +\frac{d}{2}-1\right) .
\end{equation}%
This is the condition for the creation of spin-0 bosons in a Coulomb
potential in d-dimensions, and for the three-dimensional case it reduces to 
\begin{eqnarray}
    Z\geq \allowbreak \frac{1}{e^{2}}\left( \ell +\frac{1}{2}\right) .
\end{eqnarray}
This result is analogous to that obtained in Eq.~(37) for pair creation in a magnetic monopole field \cite{Popov}.
In Table \ref{tab:week3}, we present the critical values of $Z$ for the nucleus, corresponding to different values of the quantum number, Dunkl parameter, and spatial dimension.
\begin{table}[tbph]
\centering%
\begin{tabular}{l|l|l|l|l}
\hline\hline
\rowcolor{lightgray} $d$ & $\ell _{i}$ & $Z$ ($\mu _{i}=+0.4$) & $Z$ ($\mu
_{i}=0$) & $Z$ ($\mu _{i}=-0.4$) \\ \hline
3 & 1 & $5.\,\allowbreak 7\times 137$ & $\frac{3}{2}\times 137$ & $3.3\times
137$ \\ 
4 &  & $8.6\times 137$ & $2\times 137$ & $5.4\times 137$ \\ 
5 &  & $11.5\times 137$ & $\frac{5}{2}\times 137$ & $7.5\times 137$ \\ 
6 &  & $14.4\times 137$ & $3\times 137$ & $9.6\times 137$ \\ \hline
3 & 2 & $9.7\times 137$ & $\frac{5}{2}\times 137$ & $7.3\times 137$ \\ 
4 &  & $14.6\times 137$ & $3\times 137$ & $11.4\times 137$ \\ 
5 &  & $19.5\times 137$ & $\frac{7}{2}\times 137$ & $15.5\times 137$ \\ 
6 &  & $24.4\times 137$ & $4\times 137$ & $19.6\times 137$ \\ \hline
3 & 3 & $13.7\times 137$ & $\frac{7}{2}\times 137$ & $11.3\times 137$ \\ 
4 &  & $20.6\times 137$ & $4\times 137$ & $17.4\times 137$ \\ 
5 &  & $27.5\times 137$ & $\frac{9}{2}\times 137$ & $23.5\times 137$ \\ 
6 &  & $34.4\times 137$ & $5\times 137$ & $23.6\times 137$ \\ \hline\hline
\end{tabular}%
\caption{Critical values of nucleus charge for the lowest quantum number in different spatial dimensions.}
\label{tab:week3}
\end{table}

We observe that the critical value of the nucleus charge increases with both the quantum number and the spatial dimension. Furthermore, for fixed values of the quantum number and spatial dimension, the critical value of the nuclear charge also increases as the Dunkl parameter increases.

\begin{figure}[H]
\begin{minipage}[t]{0.5\textwidth}
        \centering
        \includegraphics[width=\textwidth]{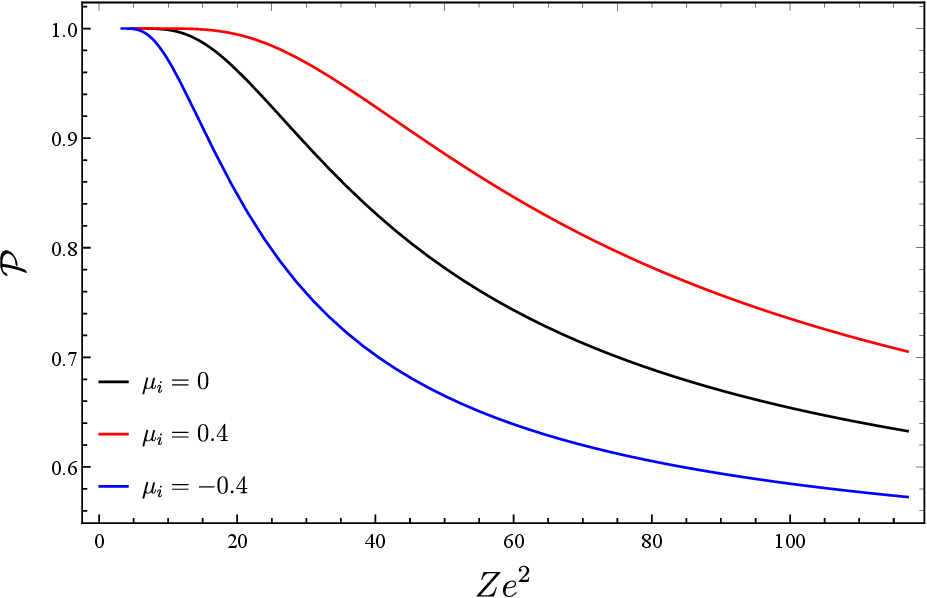}
       \label{fig:pb1}
   \end{minipage}%
\begin{minipage}[t]{0.50\textwidth}
        \centering
       \includegraphics[width=\textwidth]{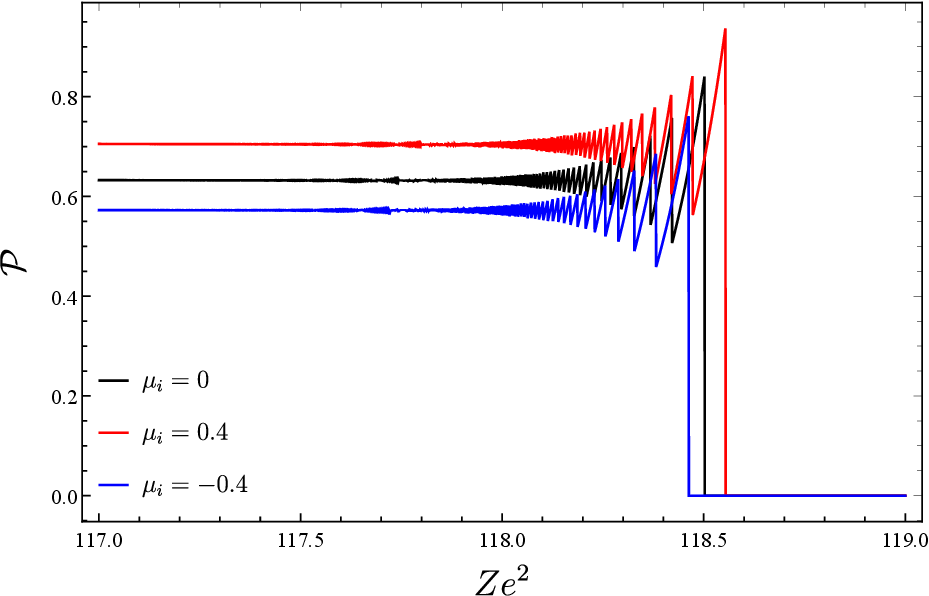}\\
        \label{fig:pb2}
    \end{minipage}\hfill
\caption{Pair creation probability as functions of $Ze^{2}$ for different values of the Dunkl parameter in three dimensions.}
\label{Efigp1}
\end{figure}

\begin{figure}[H]
\begin{minipage}[t]{0.5\textwidth}
        \centering
        \includegraphics[width=\textwidth]{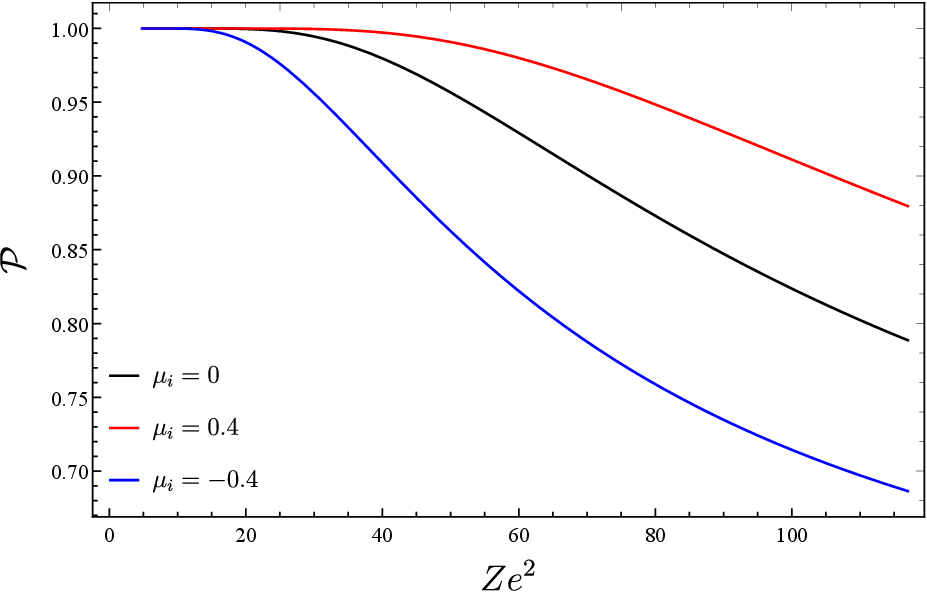}
       \label{fig:pb3}
   \end{minipage}%
\begin{minipage}[t]{0.50\textwidth}
        \centering
       \includegraphics[width=\textwidth]{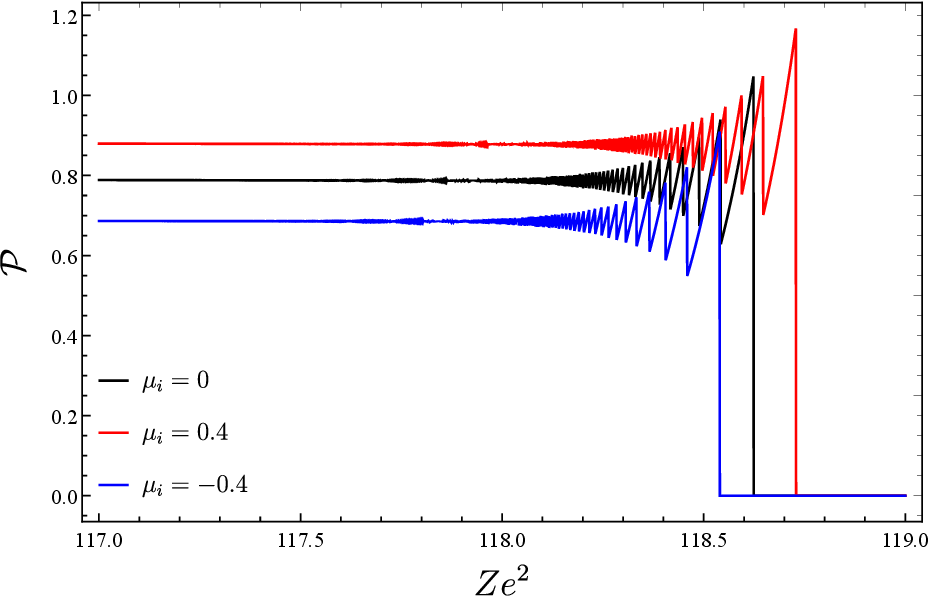}\\
        \label{fig:pb4}
    \end{minipage}\hfill
\caption{Pair creation probability as functions of $Ze^{2}$ for different values of the Dunkl parameter in four dimensions.}
\label{Efigp2}
\end{figure}

In Fig.~\ref{Efigp1} and Fig.~\ref{Efigp2}, we plot the pair creation probability curves as functions of \( Ze^{2} \) for fixed quantum number \( \ell_{i}=1 \), in both three and four dimensions, while considering the condition specified in Eq.~(\ref{cond1}).

We observe that the inclusion of Dunkl algebra increases the pair creation probability of scalar particles relative to the standard case.

\section{Conclusion}
In this paper, we extended the exact solution of the d-dimensional Klein-Gordon equation for both the harmonic oscillator and Coulomb potential cases within the framework of Dunkl algebra. We derived the d-dimensional wave functions analytically for both scenarios and analyzed the corresponding energy spectra.

For the harmonic oscillator case, we transformed the Klein-Gordon equation into a confluent hypergeometric equation through a change of variables. We determined the energy eigenvalues by applying the polynomial reduction condition to the hypergeometric function. The energy levels are expressed in terms of parameters related to Dunkl algebra, including the constant $\mu_{i}$, parities $s_{i}$, and the spatial dimension. Our results indicate that the energy eigenvalue for even parity $s_{i}=+1$ is lower than that for odd parity $s_{i}=-1$.

In the context of Dunkl algebra, we also explored the analytical solution of the Coulomb potential for both bound-state and scattering scenarios. For the bound state, we derived the exact eigensolutions and computed the energy eigenvalues. Our findings show that the energy levels are shifted and are independent of the parity $s_{i}$. In the scattering scenario, we investigated the effect of the Dunkl formalism on the pair creation rate. The solutions to the radial Klein-Gordon equation are expressed in terms of Whittaker functions. We then calculated both the probability and number density of created particles. Our results demonstrate that particle creation occurs when $Z^{2}e^{4}\geq \varpi ^{2}+\vartheta \left( \vartheta +1\right) +\frac{1}{4}$, and that Dunkl algebra enhances the number density of created spinless particles compared to the standard case.

\section*{Acknowledgments}

This work is supported by the Ministry of Higher Education and Scientific Research, Algeria under the code: B00L02UN040120230003. B. C. L. is grateful to Excellence Project PřF UHK 2211/2023-2024 for the financial support.

\end{document}